\documentclass[prb,aps,floatfix,twocolumn,amsmath,amssymb,superscriptaddress]{revtex4-1}
\usepackage{graphicx}
\usepackage{color}
\usepackage{xcolor}
\usepackage[urlcolor=blue]{hyperref}      
\hypersetup{
    colorlinks = true,                    
    citecolor = {blue},
    linkcolor = {purple},
           }
\setcitestyle{numbers,square}
\allowdisplaybreaks

\begin{document}

\title{Suppression of topological Mott-Hubbard phases by multiple charge orders in
the honeycomb extended Hubbard model}

\author{Mario Bijelic}
\email{mario.bijelic@t-online.de}
\affiliation{%
 Institute for Theoretical Physics, Goethe University Frankfurt,
 60438 Frankfurt am Main, Germany}

\author{Ryui Kaneko}
\email{rkaneko@issp.u-tokyo.ac.jp}
\affiliation{
 Institute for Solid State Physics, University of Tokyo,
 Kashiwa, Chiba 277-8581, Japan}
\affiliation{%
 Institute for Theoretical Physics, Goethe University Frankfurt,
 60438 Frankfurt am Main, Germany}

\author{Claudius Gros}
\affiliation{%
 Institute for Theoretical Physics, Goethe University Frankfurt,
 60438 Frankfurt am Main, Germany}

\author{Roser Valent\'i}
\affiliation{%
 Institute for Theoretical Physics, Goethe University Frankfurt,
 60438 Frankfurt am Main, Germany}

\begin{abstract}
We investigate the competition between charge-density-wave (CDW) 
states and a Coulomb interaction-driven topological Mott insulator
(TMI) in the  honeycomb extended Hubbard model. For the spinful 
model with on-site ($U$) and next-nearest-neighbor ($V_2$) Coulomb 
interactions at half filling, we find two peculiar six-sublattice 
charge-density-wave insulating states by using variational 
Monte Carlo simulations as well as the Hartree-Fock approximation.
We observe that conventional ordered states always win with respect
to the TMI. The ground state is given in the large-$V_2$ region
by a CDW characterized by a 220200 (001122) charge configuration 
for smaller (larger) $U$, where 0, 1, and 2 denote essentially 
empty, singly occupied, and doubly occupied sites. Within the 001122-type 
CDW phase, we find a magnetic transition driven by an emergent 
coupled-dimer antiferromagnet on an effective square lattice 
of singly occupied sites. Possible realizations of the
found states are discussed.
\end{abstract}

\maketitle

\section{Introduction}

Correlation effects in electron systems on a 
two-dimensional honeycomb lattice have been the 
subject of intensive scrutiny~\cite{castroneto2009},
both in metallic and insulating 
materials~\cite{mcchesney2010,tarruell2012}.
A special focus has been the possible realization of the quantum Hall 
effect (QHE), i.e., the quantization of the Hall conductivity in
two dimensions, not as the result of an external magnetic 
field~\cite{klitzing1980}, but due to the 
complex hopping that breaks
time-reversal symmetry~\cite{haldane1988,qi2006}.
A related
state, the quantum spin Hall (QSH), for which the spin 
Hall conductance is quantized, emerges in analogy when 
spin-orbit interactions lead to robust spin-dependent 
transport~\cite{kane2005a,bernevig2006a,bernevig2006b,koenig2007}.
The spin quantum Hall state 
has also been studied in mean-field theory for
charged-ordered triangular lattices with spin-orbit
interactions~\cite{sugita2016}.

Raghu {\it et al.}\ proposed  that a topological Mott insulator (TMI)
could be stabilized on the extended honeycomb lattice \cite{raghu2008}.
The QHE would emerge in this scenario from pure Coulomb interactions,
with an effective spin-orbit interaction being dynamically generated 
via spontaneous symmetry breaking
when the next-nearest-neighbor Coulomb
interaction is taken into account.
Here, in a spinless model, time-reversal symmetry is broken spontaneously,
while the lattice translational symmetry is preserved;
the topological phases are characterized by
the Chern number~\cite{thouless1982}.
On the other hand, for QSH in a spinful model,
$\rm SU(2)$ symmetry is broken spontaneously,
while time-reversal symmetry is preserved;
they are characterized by
the $\mathbb{Z}_2$ invariant~\cite{kane2005b}.
Nonlocal Coulomb interactions, which 
are non-negligible in graphene~\cite{wehling2011}, may also generate, however,
conventional spontaneously symmetry-broken states, such as bond 
order, charge order, and magnetic order, which then compete with
the TMI.

In the spinless Hubbard model with nearest-neighbor ($V_1$) and
next-nearest-neighbor ($V_2$) Coulomb interactions, a
Kekul\'{e} bond order phase, characterized by a $\mathbb{Z}_3$ order parameter,
has been proposed by  mean-field calculations~\cite{weeks2010}.
Exact-diagonalization (ED)~\cite{garcia-martinez2013,daghofer2014,capponi2015} and
infinite density matrix renormalization-group (iDMRG)~\cite{motruk2015}
studies also support the presence of this phase and, further, propose
much richer charge-density-wave (CDW) phases.
Possible charge instabilities have also been investigated away from half
filling~\cite{castro2011,grushin2013}.

Rich charge-density-wave phases also emerge for the spinful Hubbard
model. In addition, when the on-site Coulomb interaction $U$ is dominant,
collinear antiferromagnetic order appears~\cite{sorella2012}.
Inclusion of $V_1$ and/or $V_2$ gives rise to the competition of magnetic
and several CDW states~\cite{kurita2016,volpez2016,delapena2017}.
In this case, charge and spin degrees of freedom are intertwined, especially, off half
filling, and the formation of superstructures often enhances the geometrical
frustration of spins that allows noncollinear magnetic order to coexist
with the CDW~\cite{kaneko2016b}.

Recent extensive research on the  honeycomb extended Hubbard model
by ED~\cite{garcia-martinez2013,daghofer2014,capponi2015},
iDMRG~\cite{motruk2015},
auxiliary-field quantum Monte Carlo (AFQMC)~\cite{golor2015},
variational Monte Carlo (VMC)~\cite{kurita2016},
and functional renormalization group
(fRG)~\cite{scherer2015,volpez2016,delapena2017}
suggests that TMI is less likely than conventional ordered states, often CDW,
at half filling.
This fact leads us to ask the following questions:
\begin{itemize}
\item[(i)] Which are the plausible CDW patterns for the honeycomb extended
Hubbard model at half filling?   
\item[(ii)] Can CDW coexist with magnetic order when spin degrees of freedom
are present?
\item[(iii)] Is TMI always suppressed by CDW states? Could they coexist via the
spontaneous formation of both effective spin-orbit interaction and
charge ordering?
\end{itemize}
To answer these questions, we revisit the ground states of the extended
Hubbard model on the honeycomb lattice. We focus on the spinful model at
half filling, and consider the on-site $U$ and next-nearest-neighbor
$V_2$ Coulomb interactions, which are relevant for the TMI and CDW
states.
For simplicity, we drop the nearest-neighbor Coulomb interaction $V_1$.
We first use the restricted Hartree-Fock approximation
to get an insight into the plausible CDW states, and then apply
the VMC method with a Jastrow-Slater-type wave
function to improve the mean-field state.
We find, by the Hartree-Fock approximation,
two types of six-sublattice  CDW insulating phases:
one is a 220200-type
CDW phase realized for small $U$ and
large $V_2$, and the other is 
a 001122-type
CDW phase which appears
for larger $U$ and $V_2$.
The transition between these two is found to be of first order within
the mean-field treatment.
When the VMC method is employed, 
for a fixed large $V_2$,
we find a continuous change from the 220200-type to the
001122-type CDW phase as $U$ is increased.
The first-order transition found
in the Hartree-Fock approximation seems to melt when  quantum
fluctuations are carefully taken into account.
On the other hand, when $U$ is sufficiently large,
we again find the stable 001122-type CDW.
Within the mean-field study,
the 001122-type CDW phase shows
a magnetic transition at sufficiently
large $U$. The magnetic moment appears for two singly occupied sites
out of six sites in a unit cell, and these spins align
antiferromagnetically via superexchange interaction
(see Fig.~\ref{fig:CDW_config} for
the representation of the various phases).
Furthermore,
for the parameter region that we have studied, the energy of
the TMI state is found to always be higher than these CDW states.
The TMI does not coexist with CDW, and thus CDW is harmful for stabilizing
the TMI phase at half filling, consistent with previous studies.

\begin{figure}[t]
\includegraphics[width=0.95\columnwidth]{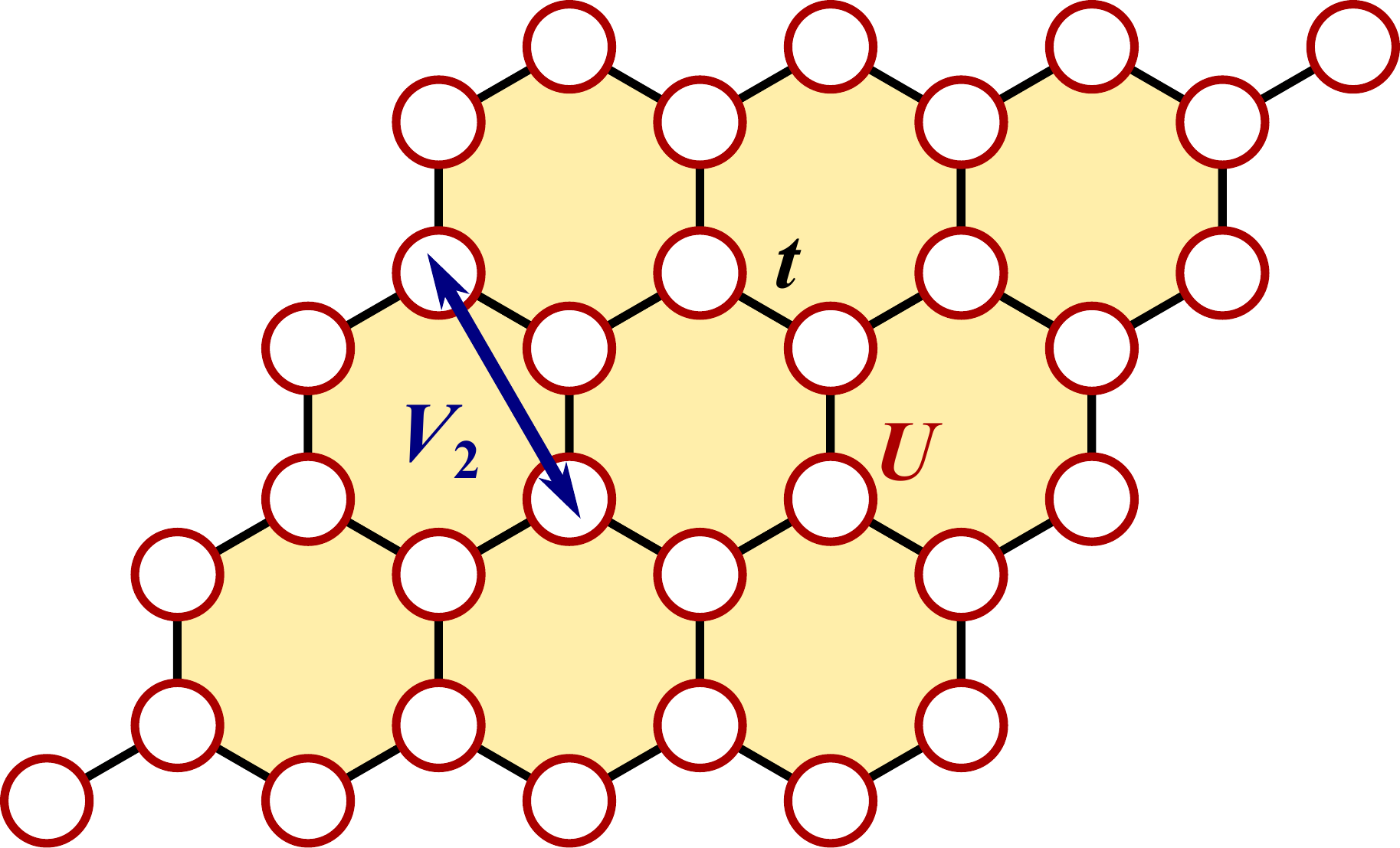}
\caption{Honeycomb lattice.
The strength of
nearest-neighbor hopping, on-site Coulomb interaction,
and next-nearest-neighbor Coulomb interaction are
denoted by $t$, $U$, and $V_2$, respectively.}
\label{fig:lattice}
\end{figure}

This paper is organized as follows:
In Sec.~\ref{sec:model_and_method},
we present the honeycomb extended Hubbard model
and introduce the Hartree-Fock approximation and the VMC method.
In Sec.~\ref{sec:results},
we present the $U$-$V_2$ phase diagram and the properties of the three
CDW phases obtained by the Hartree-Fock approximation.
We then show how quantum fluctuations modify the phases by means of the
VMC method. 
In Sec.~\ref{sec:discussion},
we comment on the origin of the charge and
magnetic orders in CDW phases.
Finally, in Sec.~\ref{sec:conclusions}, we draw our conclusions.

\section{Model and methods}
\label{sec:model_and_method}

\subsection{Extended Hubbard model}
\label{subsec:extended_hubbard_model}

We consider the extended Hubbard model on the honeycomb lattice at half filling,
\begin{eqnarray}
 H &=& -t \sum_{\langle ij\rangle \sigma}
  \left( c_{i \sigma}^{\dagger} c_{j \sigma} + {\rm h.c.} \right)
\nonumber
\\
 &&
 + U \sum_{i} n_{i\uparrow} n_{i\downarrow}
 + V_2 \sum_{\langle\langle ij\rangle\rangle} n_i n_j.
\end{eqnarray}
Here, $t$ denotes the hopping parameter and $U$ ($V_2$) denotes
the strength of on-site (next-nearest-neighbor) Coulomb interaction,
as shown in Fig.~\ref{fig:lattice}.
For simplicity, we only deal with the next-nearest-neighbor $V_2$
and neglect the nearest-neighbor $V_1$.
Hereafter, we consider  repulsive Coulomb interactions
($U, V_2\ge 0$) at half filling ($n=1$).

\subsection{Restricted Hartree-Fock method}
\label{subsec:restricted_hartree_fock_method}

\begin{figure}[t]
\includegraphics[width=0.95\columnwidth]{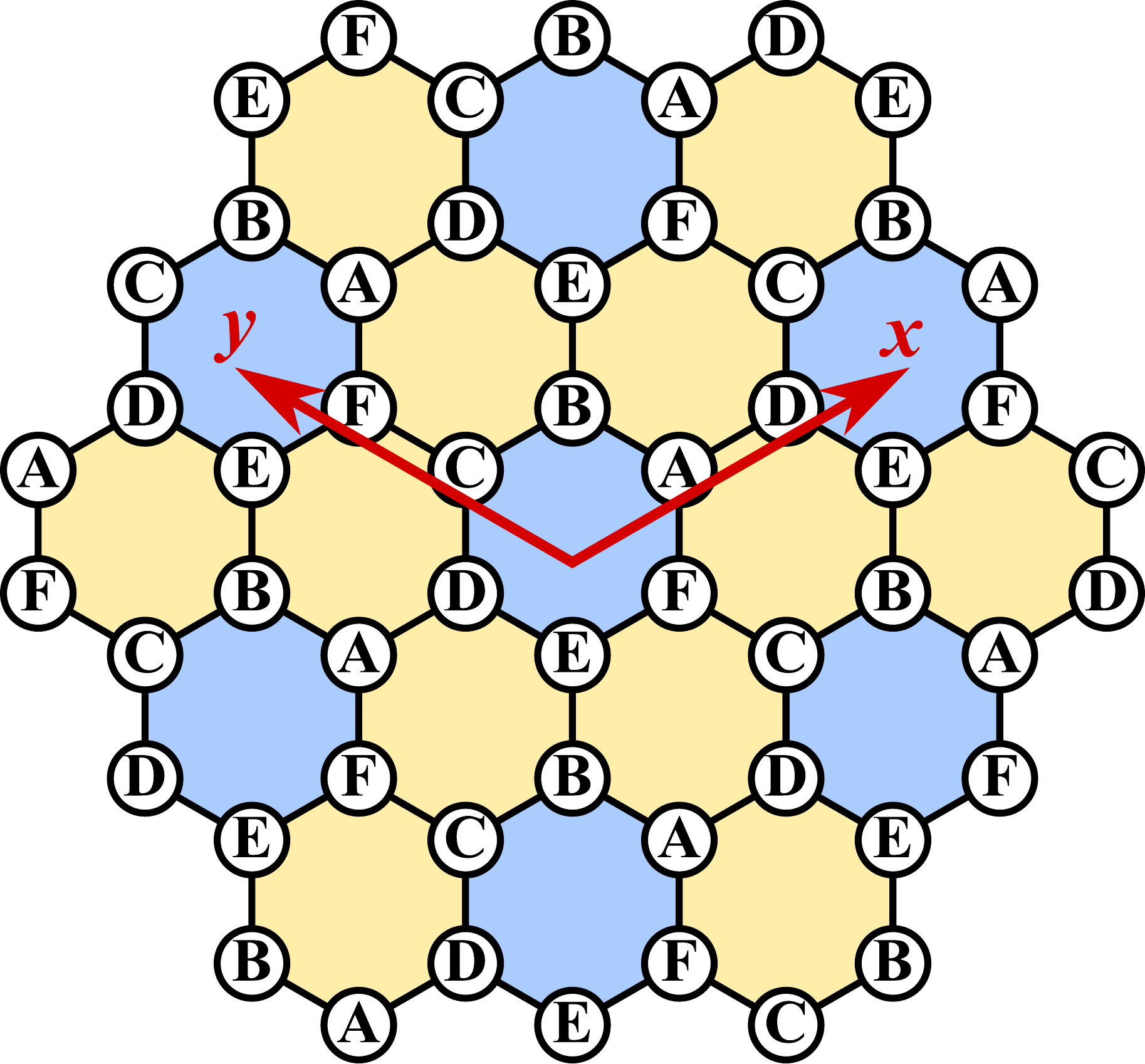}
\caption{Illustrated six independent sites
(A--F) for Hartree-Fock
mean-field calculations.}
\label{fig:6sublattice}
\end{figure}

To clarify the plausible ordered phases, we first apply the restricted
Hartree-Fock method.
To obtain the mean-field Hamiltonian,
we use the Hartree-Fock decoupling,
\begin{eqnarray}
 &&
 n_{i \sigma} n_{j \tau}
 \sim
   \langle n_{i \sigma} \rangle n_{j \tau}
 + n_{i \sigma} \langle n_{j \tau} \rangle
 - \langle n_{i \sigma} \rangle \langle n_{j \tau} \rangle
\nonumber
\\
 &&
 - \langle c_{i \sigma}^{\dagger} c_{j \tau} \rangle
   c_{j \tau}^{\dagger} c_{i \sigma}
 - c_{i \sigma}^{\dagger} c_{j \tau}
   \langle c_{j \tau}^{\dagger} c_{i \sigma} \rangle
 + \langle c_{i \sigma}^{\dagger} c_{j \tau} \rangle
   \langle c_{j \tau}^{\dagger} c_{i \sigma} \rangle,
\end{eqnarray}
where $i,j$ are site indices while $\sigma,\tau$ are spin indices,
and determine the order parameters
self-consistently.
In this paper, we consider six independent sites,
called A, B, C, D, E, and F, in a unit cell,
which give a system size of 
$N_{\rm s}=6L^2$,
with $L$ being a linear
size (see Fig.~\ref{fig:6sublattice}).

To characterize each phase, we calculate the number of electrons,
\begin{equation}
 n^{\alpha}
 =
 \frac{1}{L^2}\sum_{i\in \alpha}\left(
   \langle n_{i\uparrow} \rangle
 + \langle n_{i\downarrow} \rangle
 \right)
\end{equation}
and magnetic order parameter,
\begin{equation}
 m^{\alpha}
 =
 \frac{1}{2L^2}\sum_{i\in \alpha}\left(
   \langle n_{i\uparrow} \rangle
 - \langle n_{i\downarrow} \rangle
 \right)
\end{equation}
per each sublattice $(\alpha={\rm A,B,C,D,E,F})$.
In a charge-uniform nonmagnetic phase, $n^{\alpha}=1$ and $m^{\alpha}=0$
for all $\alpha$.
When $n^{\alpha}\not=n^{\beta}$ for a different sublattice pair of
$(\alpha,\beta)$, the phase shows charge disproportionation.
Similarly, when $m^{\alpha}\not=0$, the phase is magnetic.

In the TMI phase, the expectation value
$\langle c_{j \tau}^{\dagger} c_{i \sigma} \rangle$
gives an imaginary number for the next-nearest-neighbor-site pair $(i,j)$.
Using this value, the order parameter is defined as
\begin{eqnarray}
 \zeta
 &=&
  \frac{i}{L^2}
  \sum_{\langle\langle ij \rangle\rangle}
  \sum_{\sigma\tau}
  \nu_{ij} S_{\sigma\tau}^z
  \langle c_{j \tau}^{\dagger} c_{i \sigma} \rangle
\nonumber
\\
 &=&
 \frac{i}{2L^2}
  \sum_{\langle\langle ij \rangle\rangle}
  \nu_{ij}
  \left(
    \langle c_{j \uparrow}^{\dagger} c_{i \uparrow} \rangle
  - \langle c_{j \downarrow}^{\dagger} c_{i \downarrow} \rangle
  \right),
\end{eqnarray}
where $\nu_{ij}=-\nu_{ji}=+1(-1)$ for clockwise (anticlockwise)
orientation and $S^z=\sigma^z/2$ is a Pauli spin.
This will be a coefficient of dynamically generated spin-orbit
interaction,
and having a nonzero $\zeta$ is a necessary condition of the TMI phase.

To determine whether the phase is metallic or insulating, we calculate
the density of states and estimate the size of the charge gap.

\subsection{Variational Monte Carlo method}
\label{subsec:variational_monte_carlo_method}

To investigate the effect of quantum fluctuations beyond the mean-field
study, we employ the VMC method~\cite{robert2011} by using a Jastrow-Slater-type wave
function~\cite{gros1989} given as
\begin{equation}
 |\psi\rangle = \mathcal{J} |\phi\rangle.
\end{equation}
Here, $|\phi\rangle$ is an eigenstate of an auxiliary Hamiltonian
given by
\begin{eqnarray}
 H_{\rm aux} &=& H_{\tilde{t}} + H_{\tilde{\Delta}} + H_{\rm mag} + H_{\mu} + H_{\rm SO},
\\
 H_{\tilde{t}} &=&
  -\tilde{t} \sum_{\langle ij\rangle \sigma}
  \left( c_{i \sigma}^{\dagger} c_{j \sigma} + {\rm h.c.} \right),
\\
 H_{\tilde{\Delta}} &=&
  \tilde{\Delta}_0 \sum_{i}
  \left( c_{i \uparrow}^{\dagger} c_{i \downarrow}^{\dagger}
         + {\rm h.c.} \right)
\nonumber
\\
 &&
  + \sum_{\langle ij\rangle} \tilde{\Delta}_{ij}
  \left( c_{i \uparrow}^{\dagger} c_{j \downarrow}^{\dagger}
         + c_{j \uparrow}^{\dagger} c_{i \downarrow}^{\dagger}
         + {\rm h.c.} \right),
\\
 H_{\rm mag} &=&
  \frac{h}{2} \sum_{i} (-1)^{R_i} (n_{i\uparrow} - n_{i\downarrow}),
\\
 H_{\mu} &=&
  - \mu \sum_{i} n_i,
\\
 \mbox{and}\quad
 H_{\rm SO} &=&
  i\lambda \sum_{\langle\langle ij\rangle\rangle}
  \nu_{ij} \left(
  c_{i \uparrow}^{\dagger} c_{j \uparrow}
  - c_{i \downarrow}^{\dagger} c_{j \downarrow}
  \right),
\end{eqnarray}
where $\tilde{t}$, $\tilde{\Delta}_0$, $\tilde{\Delta}_{ij}$, $h$, $\mu$,
and $\lambda$ are variational parameters.
The honeycomb lattice contains two sites
($\rm \tilde{A}$ and $\rm \tilde{B}$) in a unit cell,
and we choose $(-1)^{R_i}=+1$ $(-1)$ for $R_i \in {\rm \tilde{A}}$
($R_i \in {\rm \tilde{B}}$).
On the other hand,
$\mathcal{J}$ is the charge Jastrow factor~\cite{jastrow1955} given by
\begin{equation}
 \mathcal{J} = \exp\left(\frac{1}{2} \sum_{i,j} v_{ij} n_i n_j\right).
\end{equation}
This wave function can represent metallic and insulating
states with and without charge disproportionation~\cite{kaneko2016a}.
We optimize the translational-invariant Jastrow factor
$v_{ij}=v(|r_j-r_i|)$ and the variational parameters in the auxiliary
Hamiltonian.
Each ordered phase is characterized by the charge order $n^{\alpha}$
and magnetic order $m^{\alpha}$ parameters, as in the Hartree-Fock
approximation.
Hereafter, we focus on the lattice systems $N_{\rm s}=6L^2$, with $L=2\sqrt{3}$ and $4\sqrt{3}$.

\begin{figure}[t]
\includegraphics[width=0.95\columnwidth]{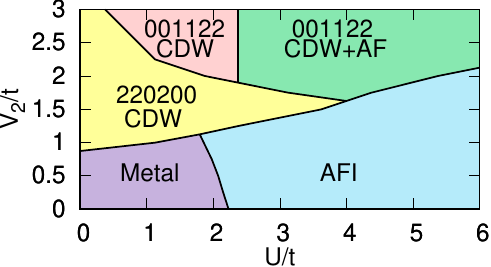}
\caption{Hartree-Fock mean-field phase diagram illustrated for
variables $V_2,\ U$ in units of the
hopping parameter $t$.
Charge and magnetic configurations of each state
are given in Fig.~\ref{fig:CDW_config}.}
\label{fig:MFPD}
\end{figure}

\begin{figure*}[t]
\includegraphics[width=0.475\columnwidth]{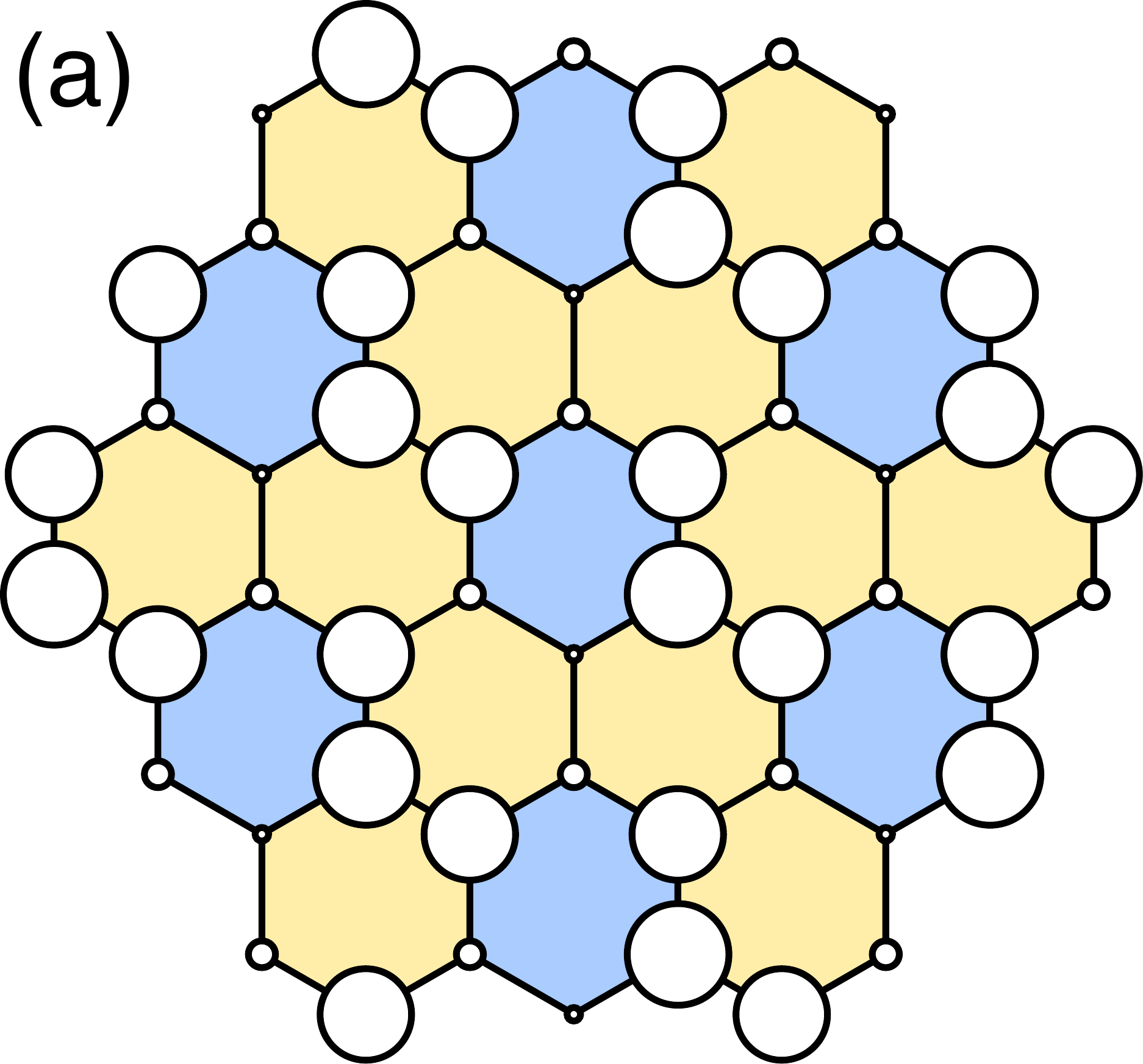}\hfil
\includegraphics[width=0.475\columnwidth]{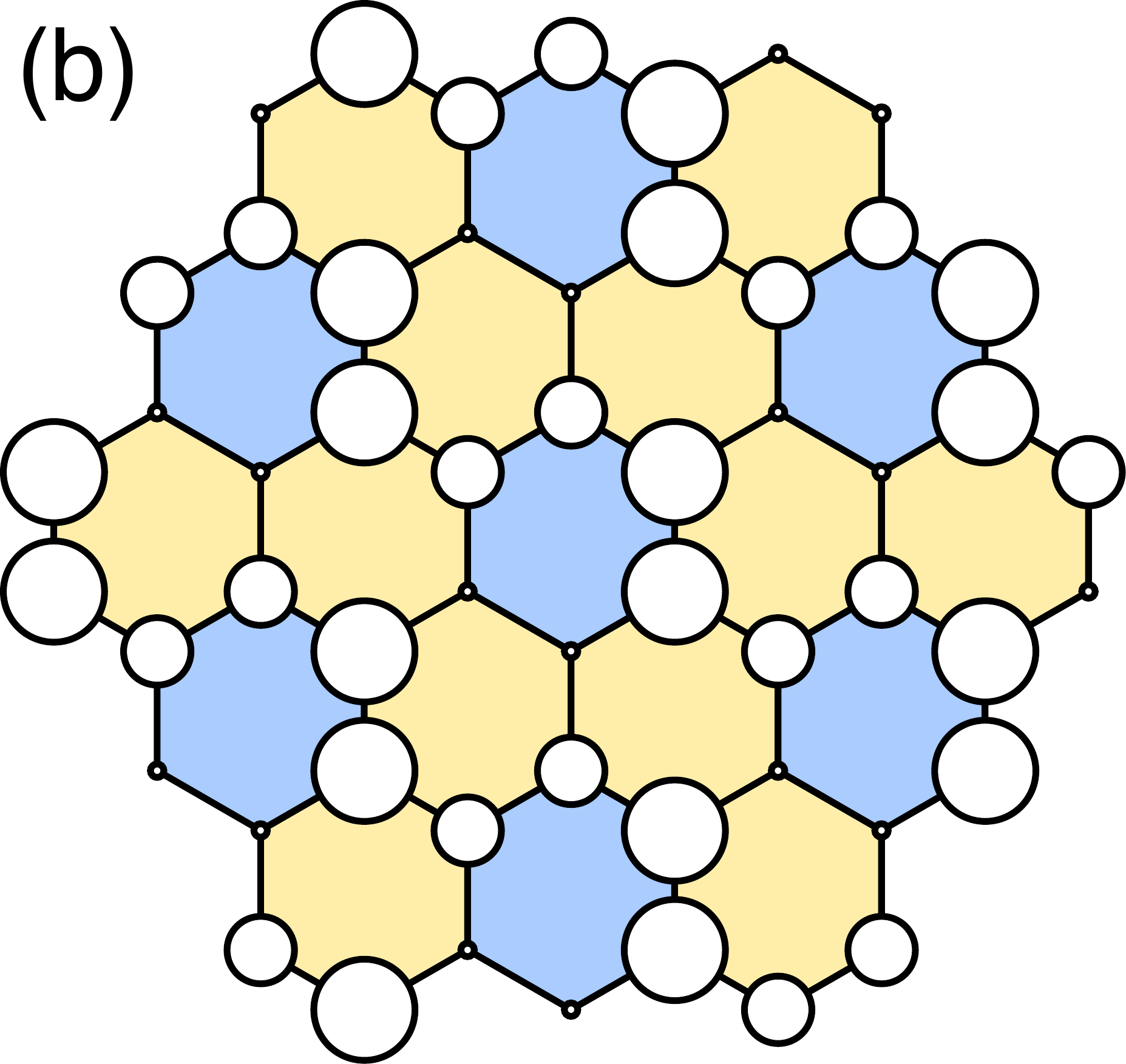}\hfil
\includegraphics[width=0.475\columnwidth]{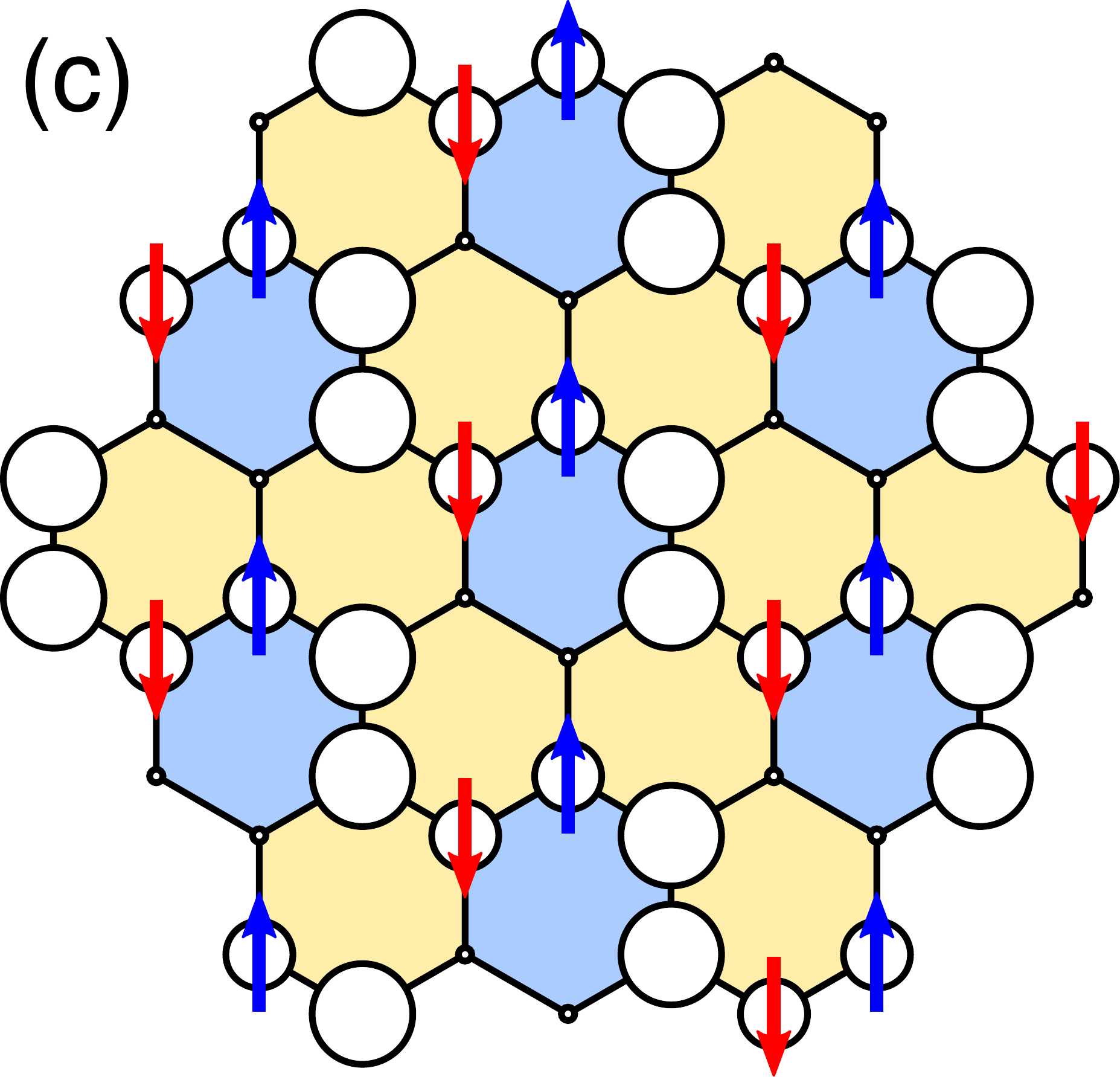}\hfil
\includegraphics[width=0.475\columnwidth]{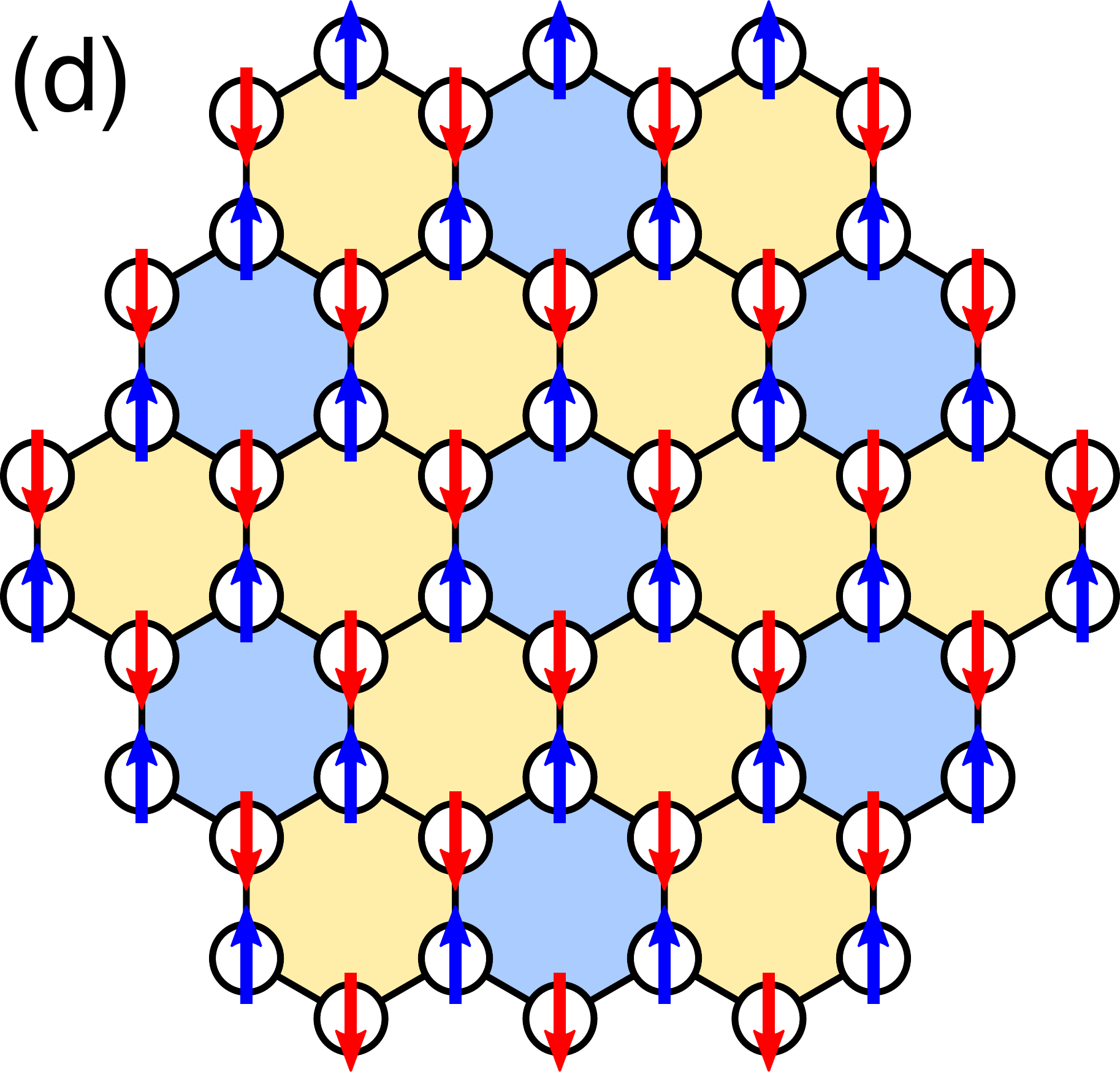}
\caption{Charge and magnetic configurations of ground states:
(a) charge configuration 220200, (b) 001122,
(c) 001122, with magnetic ordering, and (d) the AFI phase.
Each blue cell marks one unit cell.
The circle sizes denote the electron density from $0$ to $2$.
The arrows describe the spin orientation.}
\label{fig:CDW_config}
\end{figure*}

To estimate the size of the charge gap, we calculate the charge structure
factor defined as
\begin{equation}
\label{eq:Dq_sim_inv_Nq}
 N(q) = \frac{1}{N_{\rm s}} \sum_{i,j}
  \langle (n_i^{\rm \tilde{A}}+n_i^{\rm \tilde{B}})
  (n_j^{\rm \tilde{A}}+n_j^{\rm \tilde{B}}) \rangle
  e^{iq(r_i-r_j)}.
\end{equation}
In general, the charge gap $\Delta_q$ in the limit $q\rightarrow 0$ can
be estimated as~\cite{feynman1954,tocchio2011}
\begin{equation}
 \lim_{q\rightarrow 0} \Delta_q
 \propto \lim_{q\rightarrow 0} \frac{|q|^2}{N(q)}.
\end{equation}
When $N(q)\sim |q|$ for $q\sim 0$, a gap for the particle-hole
excitation vanishes and the state is metallic.
On the other hand, when $N(q)\sim q^2$ for $q\sim 0$,
$\Delta_q$ converges to a nonzero value and the state is insulating.
In practical calculations, we choose $q=q_{\rm min}=2\pi/L$.

Note that the metallic phase in the honeycomb Hubbard model shows
singular behavior $N(q)\sim |q|^2\ln|q|$ due to the presence of
a Dirac cone~\cite{sorella2016}; therefore,
the aforementioned criterion for $\Delta_q$ is 
not optimal to
distinguish metallic and insulating states.
However,
in practical VMC calculations, in most cases,
we observe an abrupt decrease of a finite-size charge gap estimated by $N(q)$,
which signals the onset of a metallic phase.
Hereafter, we first distinguish metallic and insulating phases
in a mean-field calculation, and then adopt
Eq.~(\ref{eq:Dq_sim_inv_Nq}),
as a complementary way, to estimate the size of the charge gap.

\section{Results}
\label{sec:results}

\subsection{Mean-field phase diagram}
\label{subsec:mean_field_phase_diagram}

The $U$-$V_2$ phase diagram
obtained by the restricted Hartree-Fock approximation
is shown in Fig.~\ref{fig:MFPD}.
In the absence of next-nearest-neighbor Coulomb interaction ($V_2=0$),
a continuous transition occurs at $U_c/t=2.23$
between the nonmagnetic charge-uniform semimetal and antiferromagnetic
insulating (AFI) state~\cite{sorella1992}.
These two phases remain for small $V_2$, consistent with a previous
mean-field study~\cite{kurita2016}.

On the other hand, when $V_2$ increases,
we find three insulating phases characterized by
different charge- and magnetic-order patterns.
As we detail below,
two nonmagnetic CDW phases
are similar to the ones obtained from 
mean-field~\cite{kurita2016},
fRG~\cite{volpez2016,delapena2017},
ED~\cite{garcia-martinez2013,daghofer2014,capponi2015},
and
iDMRG~\cite{motruk2015}
calculations.
In addition to the two phases,
we find a CDW phase with antiferromagnetic order
in the large-$U$ and -$V_2$ region.

\subsection{CDW states}
\label{subsec:cdw_state}

\begin{figure}[t]
\includegraphics[width=0.95\columnwidth]{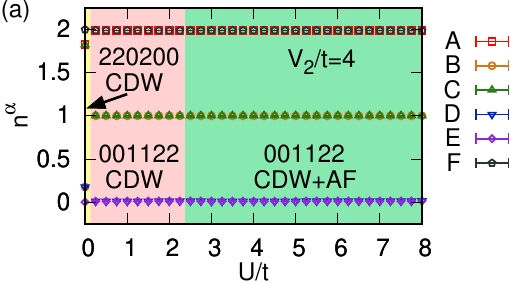}\\
\medskip
\includegraphics[width=0.95\columnwidth]{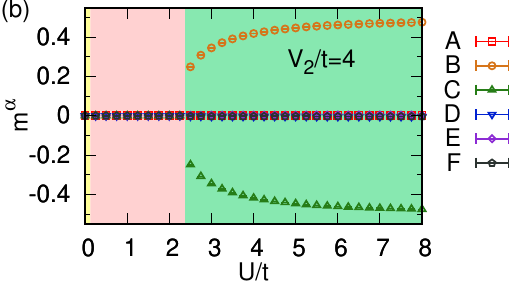}
\caption{(a) Charge and (b) magnetic distributions per lattice site
as a function of $U/t$ at $V_2/t=4$.}
\label{fig:MF_n_vs_U}
\end{figure}

\begin{figure}[t]
\includegraphics[width=0.95\columnwidth]{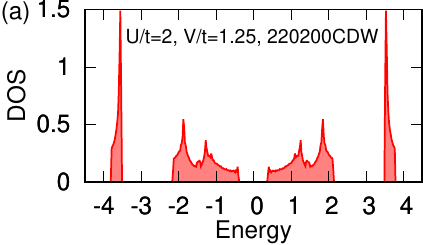}\\
\medskip
\includegraphics[width=0.95\columnwidth]{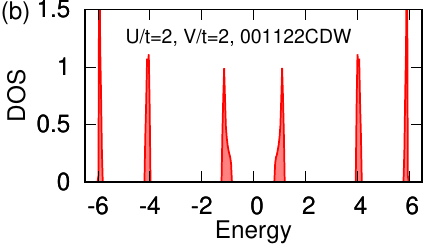}\\
\medskip
\includegraphics[width=0.95\columnwidth]{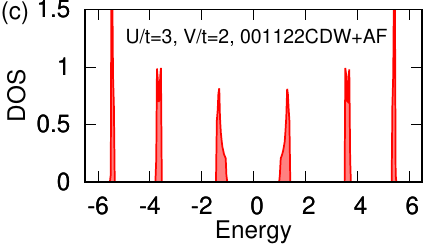}
\caption{Calculated density of states (DOS) for three CDW phases 
from the six-sublattice mean-field approximation.
A clear band gap is visible for all found charge configurations.}
\label{fig:DOS}
\end{figure}

We now discuss the charge and magnetic properties of the three CDW states
for sufficiently large $V_2$
obtained by the Hartree-Fock approximation.

In the absence of on-site Coulomb interaction ($U=0$), 
we find a six-sublattice CDW phase,
characterized by three charge-rich sites and three charge-poor sites
in a unit cell [see Fig.~\ref{fig:CDW_config}(a)].
The charge pattern is like the 220200-type,
namely, the charge-order parameters show $n^{\alpha}\sim 0$ or $2$.
However,
there is a charge disproportionation within the three charge-rich
(charge-poor) sites; one of them is richer (poorer) than the other two.
This CDW phase survives for small but nonzero $U$.

At large $U$, we find another six-sublattice CDW phase,
characterized by a 001122-type charge pattern,
namely, $n^{\alpha}\sim 0$, $1$, or $2$
[see Fig.~\ref{fig:CDW_config}(b)].

The stability of these CDW phases has been previously tested and
they were always found to stabilize in the large-$V_2$ region of  both
spinless~\cite{garcia-martinez2013,daghofer2014,capponi2015,motruk2015,
scherer2015}
and spinful models~\cite{kurita2016,volpez2016,delapena2017}.

Although  both CDW phases show six-sublattice orderings, we find a
first-order transition from one to the other.
Along a $V_2/t=4$ line,
the charge-order parameters $n^{\alpha}$ for six sites in a unit cell
show a clear jump at 
very small $U$ [see Fig.~\ref{fig:MF_n_vs_U}(a)].
However, this might be an artifact of the mean-field approximation.
As we will see later, quantum fluctuations melt the first-order transition
and give a continuous transition or a crossover.

Inside the 001122-type six-sublattice CDW phase, we find a magnetic
transition when $U$ is further increased
[see Fig.~\ref{fig:CDW_config}(c)].
Magnetic order appears for the sites with $n^{\alpha}=1$ and the
nearest spins align antiferromagnetically.
As shown in Fig.~\ref{fig:MF_n_vs_U}(b), this magnetic transition is
continuous and is distinguished from the first-order transition
between the 220200- and 001122-type CDW phases.

We also calculate the density of states to estimate the size of
the charge gap.
The number of bands are, at most, six since the unit cell contains six
sites.
A shown in Fig.~\ref{fig:DOS}, all  three CDW phases have a gap at
half filling, suggesting their insulating nature.

\subsection{Absence of a topological Mott insulator}
\label{subsec:absence_tmi}

\begin{figure}[t]
\includegraphics[width=0.95\columnwidth]{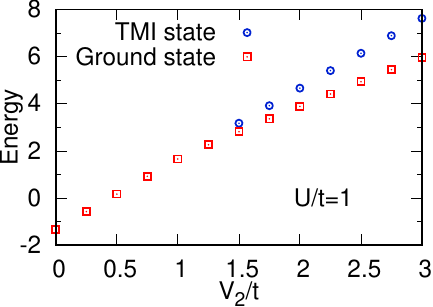}
\caption{Comparison of ground-state energies for the TMI state and the CDW state
as a function of $V_2/t$. The energy per lattice site is given in units of $t$.
The ground state (metallic phase followed by a CDW) requires a six-sublattice structure,
while the TMI state requires a two-sublattice one.}
\label{fig:ene_cmp_TMI}
\end{figure}

A topological Mott-insulating state was originally proposed for a
large-$V_2$ region~\cite{raghu2008}.
To investigate whether the TMI state can be the ground state against the
CDW states
at the Hartree-Fock mean-field level,
we compare the energy of each state,
as shown in
Fig.~\ref{fig:ene_cmp_TMI}.
The energy of the TMI state is found to always be higher than the CDW
state.

In general, the quantum Hall effect is allowed within the CDW states
for a suitable lattice structure in the presence of the spin-orbit
coupling~\cite{sugita2016}.
If such a spontaneous spin-orbit coupling is generated dynamically via
$V_2$~\cite{raghu2008}, the
coexistence of CDW and TMI is not prohibited.
However, within the parameter region we have studied,
we do not find a coexistence of CDW and TMI.
When the charge-order parameter $n^{\alpha}$ is not $1$,
we numerically find that
the expectation value
$\langle c_{j \tau}^{\dagger} c_{i \sigma} \rangle$
does not give an imaginary value,
and the order parameter $\zeta$ for the TMI is zero.

These results suggest that the TMI does not occur in the 
honeycomb extended Hubbard model at half filling.

\subsection{Variational Monte Carlo results}
\label{subsec:variational_monte_carlo_results}

\begin{figure}[t]
\includegraphics[width=0.95\columnwidth]{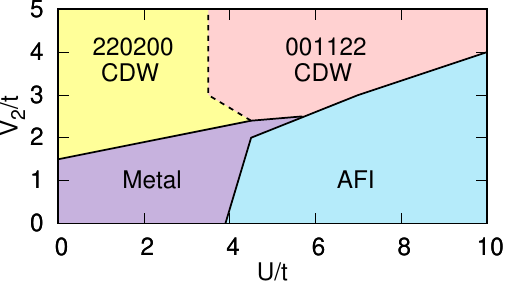}
\caption{Variational Monte Carlo phase diagram
illustrated for variables $V_2$ and $U$ in units of $t$.
Obtained phases are basically equivalent to
those given by the Hartree-Fock mean-field approximation in Fig.~\ref{fig:MFPD}.}
\label{fig:VMCPD}
\end{figure}

\begin{figure}[t]
\includegraphics[width=0.95\columnwidth]{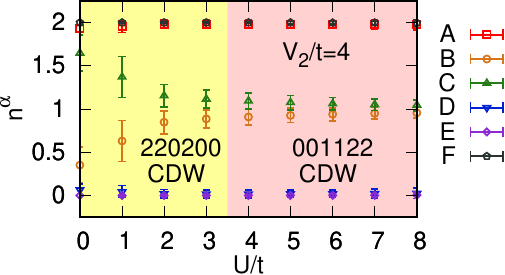}
\caption{Charge distribution per lattice site
as a function of $U/t$ at $V_2/t=4$ for $N_{\rm s}=72$
obtained by the VMC method.
The phase boundary is determined by the point where $n^c$ and $n^d$
show the same value within the statistical error bars.}
\label{fig:VMC_n_vs_U}
\end{figure}

\begin{figure}[t]
\includegraphics[width=0.95\columnwidth]{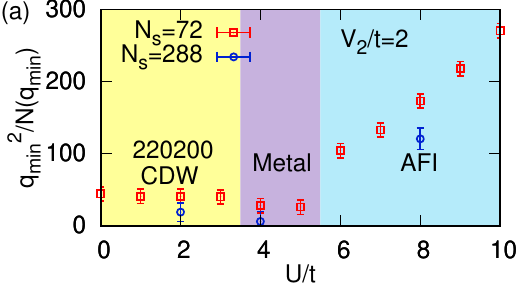}\\
\medskip
\includegraphics[width=0.95\columnwidth]{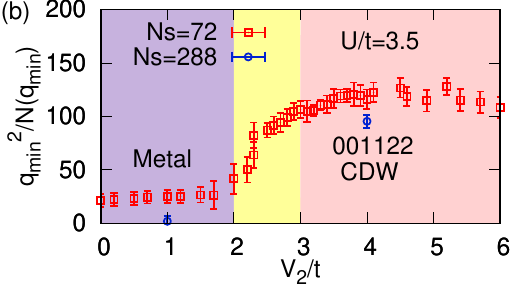}
\caption{Estimated charge gaps $q_{\rm min}^2/N(q_{\rm min})$
with $q_{\rm min}=2\pi/L$
in finite system sizes
$N_{\rm s}=6L^2$ ($L=2\sqrt{3}$ and $4\sqrt{3}$)
(a) for $V_2/t=2$ as a function of $U/t$ and
(b) for $U/t=3.5$ as a function of $V_2/t$.}
\label{fig:charge_gap}
\end{figure}

The phase diagram obtained by the VMC method is given in
Fig.~\ref{fig:VMCPD}.
As in the Hartree-Fock approximation, we find a metallic phase,
an antiferromagnetic insulating phase, and  both
220200- and 001122-type CDW phases.
For simplicity, in the present VMC study,
we do not consider possible magnetic order in the 001122-type CDW phase,
namely, the state in Fig.~\ref{fig:CDW_config}(c).

The region of validity of the ordered phases, which are overestimated in the Hartree-Fock
approximation, get shrunk
and the metallic phase region gets enlarged
when taking into account the effect of quantum
fluctuations through the Jastrow correlation factor.
Indeed, we find a single metal-insulator transition point along $V_2=0$,
estimated to be at $U_c/t\sim 3.9$, which is in good agreement with the
numerically exact value $U_c/t\sim 3.869\pm 0.013$~\cite{sorella2012}.
Furthermore, when the nearest-neighbor Coulomb interaction $V_2$ increases,
the on-site $U$ is effectively screened.
As a result, when $U$ is fixed,
a metallic phase is expected to be more stable for larger $V_2$.
Such behavior is already observed in  recent AFQMC calculations~\cite{golor2015},
and our VMC phase diagram agrees with the tendency.

Let us focus on the parameter region along $V_2/t=4$,
where the CDW phases are dominant.
The charge-order parameters $n^{\alpha}$ as a function of $U/t$ are given
in Fig.~\ref{fig:VMC_n_vs_U}.
When $U/t\sim 0$, $n^{\alpha}\sim 0$ or $2$, suggesting that the charge
pattern is 220200-like.
By contrast, when $U/t\gtrsim 4$, $n^{\alpha}\sim 0$, $1$, or $2$,
suggesting that the charge pattern is 001122-like.
The 001122 pattern is more stable for larger $V_2$.
Surprisingly, there is no clear jump in between, and these two CDW
phases are found to be continuously connected.
The first-order transition found in the Hartree-Fock approximation melts
when quantum fluctuations are carefully introduced.

A continuous change of CDW has been reported before in the
presence of $V_1$, which favors two-sublattice staggered charge
ordering~\cite{delapena2017}.
Since $V_1$ and $V_2$ favor CDW states with different
ordering wave vectors,
it is natural to expect incommensurate CDW states in the
presence of both $V_1$ and $V_2$.
However, here we
find a continuous change of CDW phases without
$V_1$.
The size of the unit cell is always kept to six sites in our
case.

When $U/t\gtrsim 10$,
charge distribution becomes uniform and the collinear
antiferromagnetic phase appears.
The transition between the CDW phase and the charge-uniform
antiferromagnetic phase is of first order, as in the Hartree-Fock
approximation.

Finally, we discuss the charge gap obtained by the VMC method.
As shown in Fig.~\ref{fig:charge_gap},
when $U$ and $V_2$ are relatively larger than $t$,
$q_{\rm min}^2/N(q_{\rm min})$ for the CDW phases
converges to a nonzero value, suggesting the two phases are insulating.
However, at smaller-$U$ values, $q_{\rm min}^2/N(q_{\rm min})$ approaches 
a tiny value, and we cannot firmly conclude that the phases are
insulating.
Larger system sizes are needed to investigate the size of the charge
gap in this region.

\section{Discussion}
\label{sec:discussion}

\subsection{Origin of charge order}
\label{subsec:origin_charge}

\begin{figure}[t]
\includegraphics[width=0.95\columnwidth]{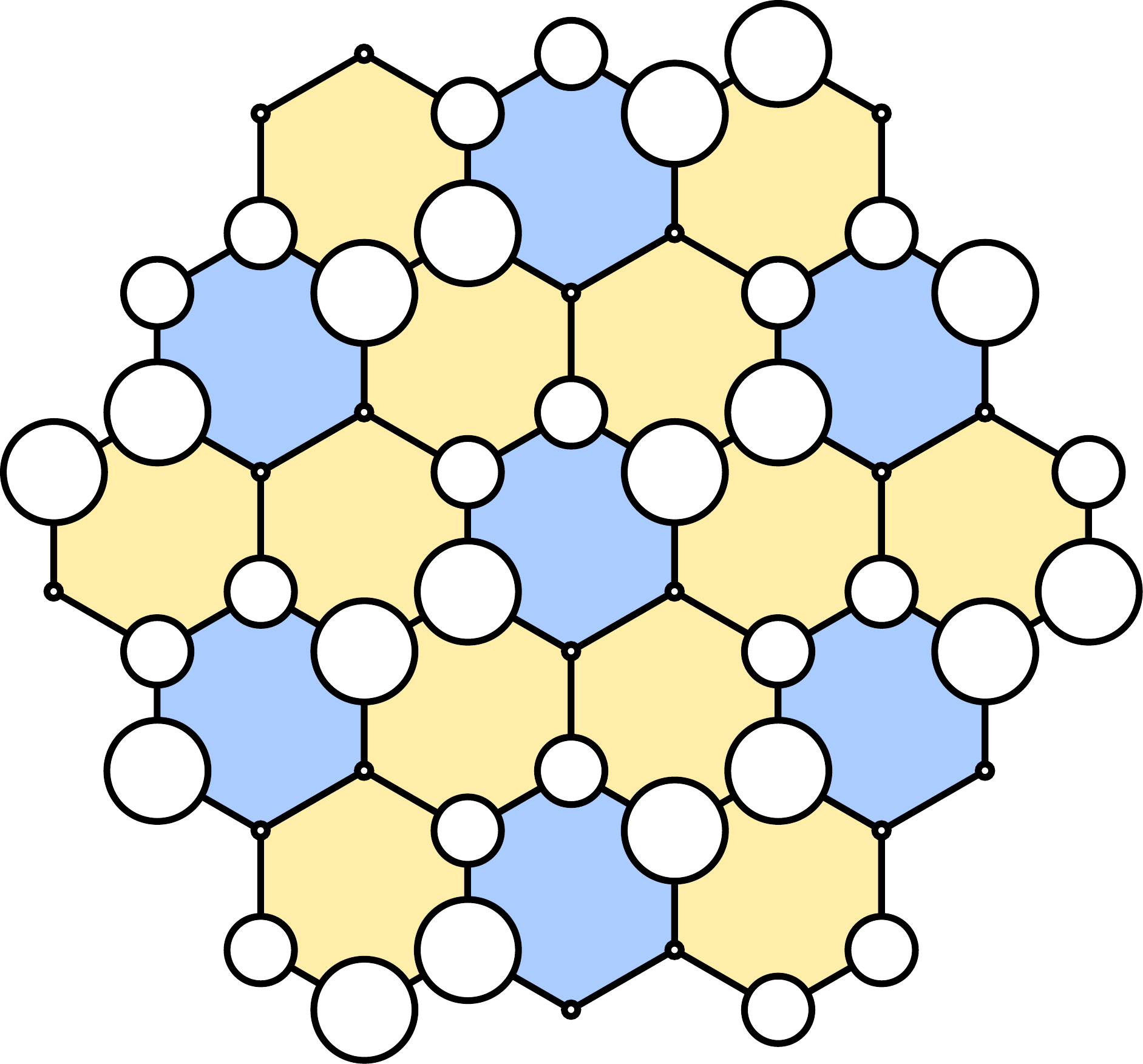}
\caption{Permuted 001122 charge configuration.
In the atomic limit ($t=0$), for $V_2\gg U$,
the energy of this configuration and
that of Fig.~\ref{fig:CDW_config}(b) are the same,
although nonzero hopping seems to choose the latter state.}
\label{fig:CDW_021120}
\end{figure}

When $V_2$ is small while $U$ is dominant,
a charge-uniform state is favorable
since $U$ prohibits double occupation of electrons.
By contrast, for the large-$U$ and -$V_2$ region,
basic charge-order patterns of the CDW phases can be
understood by considering the atomic limit ($t=0$).

Let us first consider the case of small but nonzero $U$.
When $t=0$ while $V_2\gg U$ is turned on,
the original honeycomb lattice decouples into
doubled triangular lattices.
One has only to minimize the energy for each sublattice $\tilde{A}$ or
$\tilde{B}$, separately, by choosing the plausible charge-ordering
patterns.
It is enough to find such configurations
by minimizing
the energy of a local triangle connected by $V_2$.
At half filling, the number of electrons per triangle should be three.
Besides,
positive $V_2$ favors empty sites to reduce the energy loss in
$V_2 n_i n_j$, while on-site $U$ disfavors doubly occupied sites.
Consequently,
the optimal charge configuration is a 012-type charge pattern.
Note that the number of degenerate configurations is finite,
but still large.
The 012 pattern is sixfold degenerate for each sublattice
$\tilde{A}$ or $\tilde{B}$, and, therefore, the total number of degeneracy
on the honeycomb lattice is $36$.

These configurations can be divided into two types
that cannot be transformed to each other by mirroring or rotation;
namely, 001122-type [see Fig.~\ref{fig:CDW_config}(b)]
and 021120-type [see Fig.~\ref{fig:CDW_021120}]
charge orderings.
Both of them are 18-fold degenerate, respectively.
In the atomic limit, these two states have exactly the same energy;
however, quantum fluctuations lift the degeneracy.
The Hartree-Fock approximation predicts that
nonzero hopping $t$ favors the 001122-type pattern.

In contrast to the $U>0$ case, in the absence of $U$, 
the ground state in the atomic limit shows macroscopic degeneracy.
Since double occupation is no longer prohibited by $U$,
the number of singly occupied sites is flexible.
Meanwhile, the total number of electrons should be identical to the
number of sites at half filling.
Therefore, any two different local charge patterns $[(012),(012)]$
can be turned into $[(002),(022)]$
without any additional energy cost.
As in the $U>0$ case, nonzero hopping $t$ lifts the degeneracy
and seems to select the 220200-type CDW phase.

When $U\gg t$, the transition point between
the charge-uniform antiferromagnetic collinear state
and 001122-type CDW state
can be estimated by comparing their energies.
On a single triangle connected by $V_2$,
the charge-uniform state gives $E=3V_2$ per site when the spins are assumed to be
fully polarized, while the CDW state gives $E=2V_2+U/3$ per site.
As a result, the CDW phase is stable when $3V_2\ge U$.
The phase boundaries obtained by the Hartree-Fock approximation
and the VMC method are in good
agreement with this prediction.

\subsection{Origin of magnetic order}
\label{subsec:origin_magnet}

\begin{figure}[t]
\includegraphics[width=0.95\columnwidth]{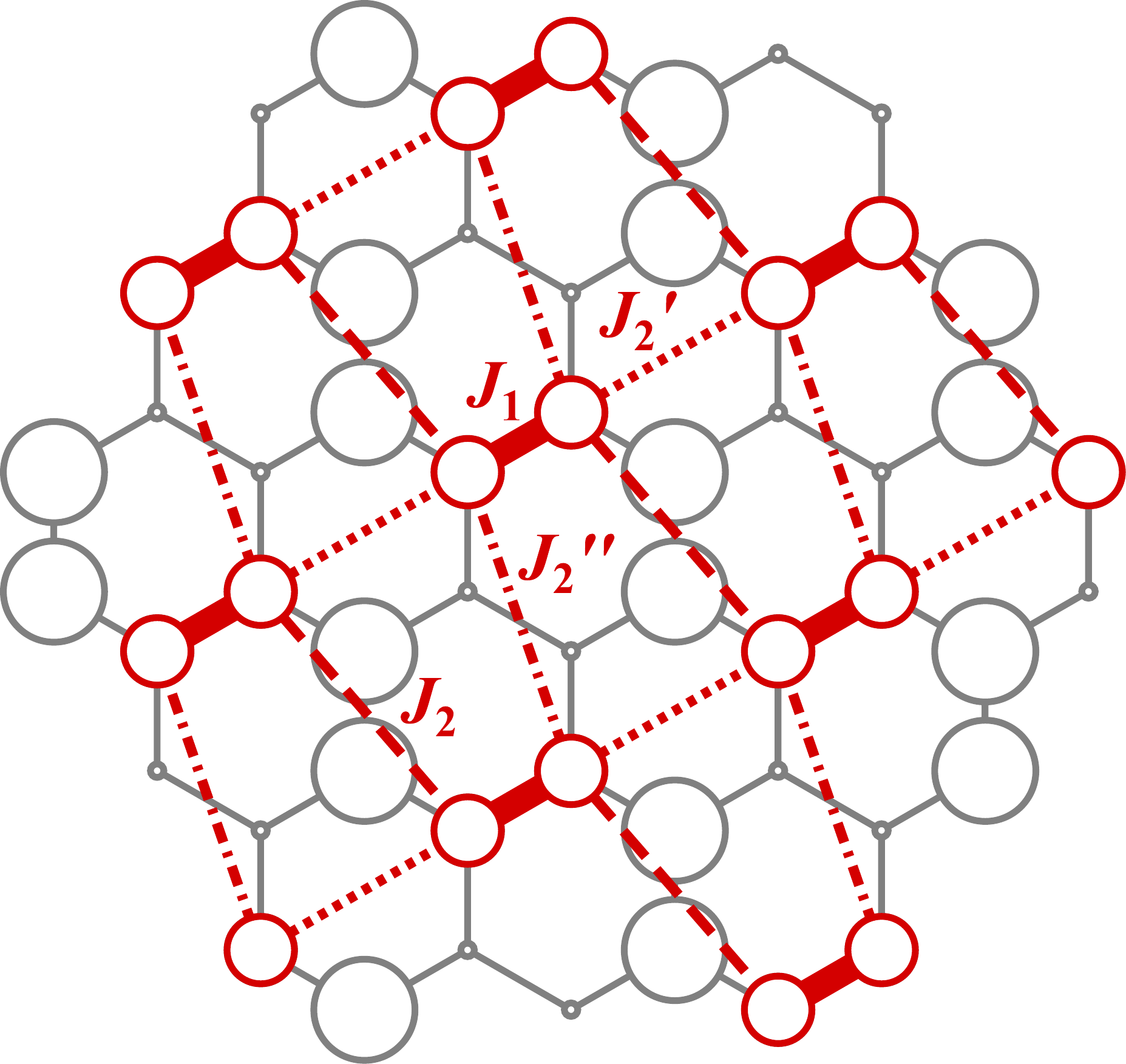}
\caption{Effective $J_1{}-{}J_2{}-{}J_2'-{}J_2''$ square Heisenberg model
emerged from the 001122-type CDW.
The antiferromagnetic N\'{e}el order is expected for $J_1\sim J_2,J_2',J_2''$,
while the dimer order is expected for $J_1\gg J_2,J_2',J_2''$.
The sizes of $J_1$, $J_2$, $J_2'$, and $J_2''$ are determined by
the strength of $U/t$ and $V_2/t$.}
\label{fig:j1_j2_model}
\end{figure}

Let us consider the origin of magnetic order coexisting with CDW for
large $U$ and $V_2$.
In the 001122-type CDW phase, spin degrees of freedom survive for
singly occupied sites [see Fig.~\ref{fig:CDW_config}(b)], and
superexchange interaction $J$ is generated via virtual hopping
processes.
For the singly occupied sites next to each other,
$J_1$ is generated by the second-order process.
Similarly,
for the singly occupied sites connected by the $r=3$ Manhattan distance,
$J_2$, $J_2'$, and $J_2''$ are generated by the sixth-order process.
These processes realize the
$J_1{}-{}J_2{}-{}J_2'-{}J_2''$ square Heisenberg model
with $J_1\ge J_2, J_2', J_2''$,
as shown in Fig.~\ref{fig:j1_j2_model}.

When $J_1\sim J_2,J_2',J_2''$,
this is basically the simple antiferromagnetic square Heisenberg
model, 
and conventional antiferromagnetic N\'{e}el order is expected.
On the other hand,
when $J_1\gg J_2,J_2',J_2''$,
the system nearly decouples into isolated dimers,
and staggered dimer order is expected.
Such a coupled-dimer antiferromagnetic
model~\cite{matsumoto2001,wenzel2008,jiang2012,yasuda2013}
has been  previously numerically studied.
For $J_2=J_2'=J_2''$, the N\'{e}el-dimer transition occurs at
$J_2/J_1\sim 0.4$~\cite{wenzel2008,jiang2012,yasuda2013}.

In the simple spin system,
the nature of this N\'{e}el-dimer transition is believed to belong to
the three-dimensional classical Heisenberg
universality~\cite{jiang2012,yasuda2013},
although its numerical detection is known to be difficult
due to the large finite-size effect~\cite{wenzel2008,jiang2012,yasuda2013}.
One may expect the same scenario in the present system;
however, spin and charge degrees of freedom are coupled in the Hubbard model.
Furthermore, the anisotropic spin-exchange interactions
are not static, but generated by spontaneous charge order.
These factors possibly modify the nature of the transition
and make it much harder to identify the magnetic transition numerically.

The Hartree-Fock approximation seems to find 
these magnetic and nonmagnetic CDW states
although the mean-field method, in principle, cannot represent
dimer singlet pairs.
Besides, the effective ratio $J_2/J_1$ could be a nonmonotonic
function of $U$ and $V_2$. Therefore,
it is not trivial how dimer and magnetic phases appear
in the $U$-$V_2$ phase diagram.
Quantum fluctuations may modify the 
mean-field magnetic-nonmagnetic phase boundary
significantly inside the CDW phase.
Investigating the N\'{e}el-dimer transition point and its nature is beyond
the scope of this paper, and it is left as a  subject for future study.

\subsection{Possible realizations}

In general, Coulomb interactions become smaller as the distance is increased,
and hence it is not simple to realize $V_1<V_2$.
However, these CDW phases may be found in 
silicon adatoms~\cite{tosatti1974,carpinelli1996}
or
optical lattices~\cite{tarruell2012}
when the lattice forms a double-layered triangular structure.
This is because
the next-nearest-neighbor Coulomb interaction $V_2$ on the original
honeycomb lattice corresponds to the nearest-neighbor one in
triangular lattices, and $V_2$ could be dominant when the layers are far enough.

\section{Conclusions}
\label{sec:conclusions}

We have investigated the phase diagram of the Hubbard model on the honeycomb lattice
at half filling in the presence of on-site ($U$) and next-nearest-neighbor ($V_2$)
Coulomb interactions.
By applying the restricted Hartree-Fock approximation,
 we find three six-sublattice CDW insulating phases, namely,
(i) a nonmagnetic 220200-type CDW phase,
(ii) a nonmagnetic 001122-type CDW phase,
and
(iii) a magnetic 001122-type CDW phase,
as well as the semimetal and antiferromagnetic insulating phases.
To investigate the stability of CDW phases beyond the mean-field study,
we further
apply the VMC method with Jastrow-Slater-type wave functions.
We find that  quantum fluctuations destroy the first-order
transition found in the mean-field approximation,
and give a continuous change from 220200-type CDW to
001122-type CDW.
On the other hand, when $U$ and $V_2(\ge U/3)$ are large enough, we find
a solid 001122-type CDW phase.

In contrast to previous studies on the spinful model,
where nonmagnetic CDW phases were proposed,
we find a magnetic transition within the CDW phase.
In the magnetic CDW phase,
only the singly occupied sites contribute to magnetism,
and they show collinear antiferromagnetic order.
To understand the origin of magnetic order,
we considered the possible superexchange interactions through virtual
hopping processes.
The spins in the 001122-type CDW phase are found to behave as in
a coupled-dimer antiferromagnetic model on the square lattice,
where nonmagnetic and magnetic phases are determined by the anisotropy
of the spin-exchange interaction.

For the parameter region we have studied,
the TMI phase is found to be less favorable than
the CDW phases, and there is no coexisting region of TMI and CDW.
Therefore, CDW is harmful for the TMI phase
in the honeycomb extended Hubbard model at half filling.
These CDW phases may be found in silicon adatoms or optical lattices
where sizable $V_2$ on the honeycomb lattice could be realized
by favoring a double-layered triangular structure.

\begin{acknowledgments}
M.B., R.K.,
and R.V.\ acknowledge the support of the German Science Foundation (DFG)
through Grant No.\ SFB/TRR49.
\end{acknowledgments}


%

\end{document}